\documentclass[preprint,journal]{vgtc}       

\ifpdf
  \pdfoutput=1\relax                   
  \usepackage{graphicx}                
  \DeclareGraphicsExtensions{.pdf,.png,.jpg,.jpeg} 
\else
  \usepackage{graphicx}                
  \DeclareGraphicsExtensions{.eps}     
\fi%

\graphicspath{{figures/}{pictures/}{images/}{./}} 

\usepackage{microtype}                 
\PassOptionsToPackage{warn}{textcomp}  
\usepackage{textcomp}                  
\usepackage{mathptmx}                  
\usepackage{times}                     
\usepackage{cite}                      
\usepackage{tabu}                      
\usepackage{booktabs}                  

\usepackage[table]{xcolor}
\usepackage{graphics}
\usepackage{amssymb}
\usepackage{enumitem}
\usepackage{microtype}

\definecolor{c1}{rgb}{0,0.3,1}
\definecolor{c2}{rgb}{1,0,0.0}
\definecolor{c3}{rgb}{0.16, 0.5, 0.0}
\definecolor{c4}{rgb}{0.2, 0.41, 0.65} 
\definecolor{c5}{rgb}{0.1, 0.65, 0.48}
\definecolor{c6}{rgb}{1,0.5,0}
\definecolor{row1}{HTML}{4285F4}
\definecolor{row2}{HTML}{FF6D01}
\definecolor{row3}{HTML}{34A853}

\newcolumntype{?}{!{\vrule width 1pt}}

\onlineid{1019}

\vgtccategory{Research}
\vgtcpapertype{application/design study}

\title{Motif-Based Visual Analysis of Dynamic Networks}

\author{Eren Cakmak, Johannes Fuchs, Dominik J\"ackle, Tobias Schreck, Ulrik Brandes, and Daniel Keim}
\authorfooter{
\item Eren Cakmak, Johannes Fuchs, and Daniel Keim are with the University of Konstanz, Germany. E-mail: firstname.lastname@uni-konstanz.de
\item Dominik J\"ackle is an Independent Researcher, Munich, Germany. \\ E-Mail: dominikjaeckle@gmail.com
\item Tobias Schreck is with the TU Graz, Austria. \\ E-Mail: tobias.schreck@cgv.tugraz.at
\item Ulrik Brandes is with the ETH Z\"urich, Switzerland. \\ E-Mail: ubrandes@ethz.ch
}

\shortauthortitle{Cakmak \MakeLowercase{\textit{et al.}}: Motif-Based Visual Analysis of Dynamic Networks}

\abstract{Many data analysis problems rely on dynamic networks, such as social or communication network analyses.
Providing a scalable overview of long sequences of such dynamic networks remains challenging due to the underlying large-scale data containing elusive topological changes.
We propose two complementary pixel-based visualizations, which reflect occurrences of selected sub-networks (motifs) and provide a time-scalable overview of dynamic networks: 
a \textit{network-level census} (motif significance profiles) linked with a \textit{node-level sub-network metric} (graphlet degree vectors) views to reveal structural changes, trends, states, and outliers.
The network census captures significantly occurring motifs compared to their expected occurrences in random networks and exposes structural changes in a dynamic network. 
The sub-network metrics display the local topological neighborhood of a node in a single network belonging to the dynamic network.
The linked pixel-based visualizations allow exploring motifs in different-sized networks to analyze the changing structures within and across dynamic networks, for instance, to visually analyze the shape and rate of changes in the network topology.
We describe the identification of visual patterns, also considering different reordering strategies to emphasize visual patterns.
We demonstrate the approach's usefulness by a use case analysis based on real-world large-scale dynamic networks, such as the evolving social networks of Reddit or Facebook.
} 


\CCScatlist{ 
 \CCScat{K.6.1}{Management of Computing and Information Systems}%
{Project and People Management}{Life Cycle};
 \CCScat{K.7.m}{The Computing Profession}{Miscellaneous}{Ethics}
}

\teaser{
   \centering
   \includegraphics[width=\linewidth]{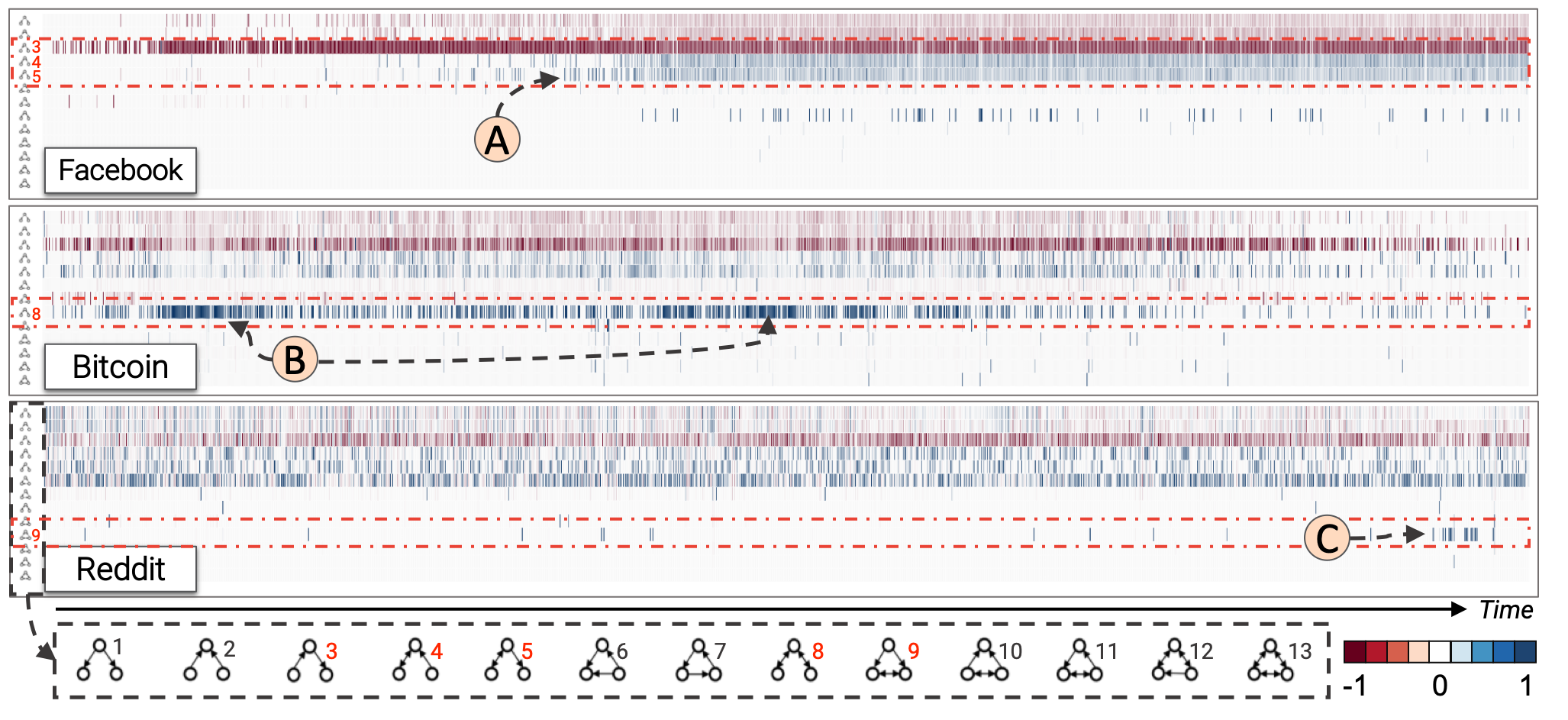}
   \caption{
    The network-level census (motif significance profile) view provides a scalable overview of evolving sub-networks (motifs) by presenting every day as a colored pixel bar.
    The three visualizations show distinct real-world dynamic networks.
    The x-axis displays the temporal dimension, and the y-axis depicts the network census of the thirteen triad motifs. 
    The divergent color map contrasts under-represented (red) and over-represented (blue) motifs to their expected occurrences in randomly generated networks. 
    (A) highlights a significant structural change between two temporal states in a social network. 
    The 4 and 5-triads reflect the mutual posting and replying behavior between friends in the evolving social network. 
    (B) emphasizes two temporal trends with 8-triads representing the behavior that two users give each other trust ratings after transactions. 
    (C) indicates an anomaly period in which the 9-triad is over-represented due to NFL Superbowl.
    Section~\ref{sec:use-case-2} details the datasets and labels (A-C).}
  \label{fig:use-case-2-1}
 }

\vgtcinsertpkg

\begin{document}


\firstsection{Introduction}

\maketitle

\firstsection{Introduction}\label{sec:intro}
\maketitle 
Many data exploration and analysis problems rely on network representations in one form or another. 
Besides the visual analysis of static networks, i.e., where nodes and relationships are fixed, in many domains \emph{dynamic} network data arises. 
These, in turn, pose challenging questions about the change of the network structure, features, and patterns over time.
For example, social networks, computer networks, or transportation networks depend on and change over time. While first examples of visual analysis of dynamic networks have recently explored real-world applications~\cite{beck2017taxonomy}, such as in biology~\cite{hadlak2011situ} or communication analyses~\cite{hadlak2013supporting}, it remains a challenging problem.

A typical user task in such applications is to obtain an overview by identifying similar and dissimilar network structures over time to understasnd topological changes~\cite{van2016reducing}.
However, providing a scalable overview of changing network structures remains challenging due to large-scale network data that usually evolve over long periods. 
For instance, the growing linkage behavior of the social network Reddit~\cite{kumar2018community} consists of five years of data with roughly 55K nodes and 850K edges.
Previous dynamic network visualizations, therefore, regularly utilize abstraction methods to reduce the complexity and provide a high-level overview of temporal changes~\cite{van2016reducing}.
However, such abstraction methods depend on the graph size, the frequency of changes, and the extracted global or local metrics (e.g., diameter or node degrees).
A promising approach is a local analysis of sub-networks (e.g., motifs or graphlets) that define and provide insight into complex network topologies~\cite{vazquez2004topological}.
However, in visualization research, sub-networks are mainly used to abstract and display static networks. 
For instance, Dunne and Shneiderman~\cite{dunne2013motif} display motifs as simplified glyph representations.
To this day, dynamic network visualizations refrained from using motifs or graphlets, although they can provide useful, scalable overviews of evolving sub-networks. 

{\looseness=-1}
In this work, we propose two complementary scalable pixel visualizations~\cite{keim2000designing} to provide an overview of changing motif structures in large-scale dynamic networks.
The first pixel-based representation of \textit{network-level census} displays significantly occurring motifs to reveal structural changes, trends, states, and outliers.
The visualization allows users to compare topological structures within and across several dynamic networks.
Moreover, we propose a second linked pixel-based representation of \textit{node-level sub-network metrics} that presents detailed node neighborhood information and allows us to compare individual networks within a dynamic network in more detail.
We introduce potential visual patterns and discuss different reordering strategies to emphasize visual patterns, for instance, rearranging the pixel representations based on network metrics to highlight similar network superfamilies.
We likewise display a node-link diagram juxtaposed to relate the visual patterns with the network topology. 
To improve the scalability of our approach, we apply clustering to find superfamilies of similar networks and node neighborhoods.
Both pixel-based visualizations allow identifying, comparing, tracing, and interpreting similar evolving network topologies to understand evolving sub-network structures in dynamic networks.
We demonstrate the usefulness of our approach through three use cases analyzing synthetic and real-world datasets.

Our  main contributions are:
(1) we discuss and exploit the possibilities of a motif analysis to provide an overview of significant structural changes in dynamic networks, 
(2) we visualize the results with two linked pixel visualizations and discuss the applicability of reordering strategies to emphasize visual patterns, 
and (3) we implement a prototype to evaluate the usefulness of our approach in a use case analysis with real-world data.
\section{Background \& Related Work} \label{sec:related_work}
Next, we first provide background information and discuss related motif-based visualization approaches.
The discussed research is selected based on the surveys of Kerracher et al.~\cite{kerracher2015task}, Borgo et al.~\cite{borgo2015the}, Beck et al.~\cite{beck2017taxonomy}, Nobre et al.~\cite{nobre2019state}, and Ribeiro et al.~\cite{ribeiro2021survey}.
Finally, we compare and delineate our work from related approaches.

\subsection{Background}
Dynamic networks model evolving relationships between real-world entities in various application domains, such as social network analysis.
A dynamic network $DN$ can be defined as a series of $T$ static graphs $DN = (N_1,N_2,...,N_T )$. 
Where each network $N_i = (V_i, E_i)$ at the time step $i$ consists of a set of vertices or nodes $V_i$ and a set of directed edges $E_i \subseteq V_i \times V_i$.
In our work, we follow the common visualization terminology and use the term network to describe graphs in which nodes and edges have attributes~\cite{von2011visual}.
Next, since our approach analyzes motifs, network census, and graphlets, we briefly introduce these terms.

\textbf{Motifs} are regarded as the basic building blocks of a network and are frequently defined as significantly over-represented directed sub-networks~\cite{milo2002network}.
Network motifs are crucial in various application domains to analyze topological structures, such as in co-authorship networks~\cite{choobdar2012comparison} or brain networks~\cite{sporns2004motifs}.
Generally, a network motif is a distinct sub-network that occurs more often than expected in a random reference network model~\cite{milo2002network}. 
Likewise, motifs that are significantly under-represented are considered to be anti-motifs.
In this work, we consider motifs as induced sub-networks, meaning that all existing edges between nodes are always included in the sub-network.

\textbf{Network censuses} (motif significance profiles) help to identify and compare different-sized networks ~\cite{milo2004superfamilies}.
The census is computed by counting the specific number of motifs $m_i$ in a network $N^{real}$ and normalizing occurrences of the motifs to a set of randomized networks $N^{rand}$ with the same degree sequence. 
The statistical significance is defined as $Z_i = (N_{i}^{real} - N_{i}^{rand}) / std(N_{i}^{rand})$ with $N_{i}^{real}$ being the real number of motif occurences, and $N_{i}^{rand}$ the occurrence of the motif in a randomized network. 
By default, we use the configuration model~\cite[Chapter~4]{newman2003structure} as a null model to create random networks for the computation of the network census.
The normalized significance profile is defined as follows: $SP_i = Z_i / \sqrt{\sum_{j}{Z_{j}^{2}}}$.
$SP_i$ indicates the relative significance of the motif $m_i$ compared to the frequency of the same motif $m_i$ in a randomly generated network with the same degree sequence.
The values of $SP_i$ are between $[-1,1]$, with $SP_i = 1$ indicating that the motif is significantly over-represented, and inversely $SP_i = -1$ defines an anti-motif, meaning that the motif is significantly under-represented.
Furthermore, thirteen triad motifs with three nodes and without self-loops are often used to compute the network census~\cite{milo2004superfamilies}.
We also facilitate these triads since the motifs capture the lowest level of social structures, considering relations between three nodes~\cite{holland1977method}.
The triads are crucial for studying social networks, such as triadic closures or transitivity~\cite{lou2013learning}.
Such triads are also mainly used in statistical models for dynamic networks, such as the stochastic actor-oriented models~\cite{snijders2001statistical} and temporal exponential random graph models~\cite{hanneke2010discrete}.

\textbf{Graphlets} are non-isomorphic induced undirected sub-networks without the concept of significance and over-representation~\cite{prvzulj2007biological}.
Graphlets can be used to calculate the topological similarity between nodes from different networks~\cite{milenkovic2010optimal}.
Graphlets are connected sub-networks and capture the instances of induced motifs occurring in the neighborhood of a node.
The graphlets for two to five nodes around a particular node are known as the 73-dimensional graphlet degree vector (GDV)~\cite{hayes2013graphlet}.
GDVs enable us to compare the topological similarity between nodes and are essentially the neighborhood signature of four hops around a given node.
There are various graphlet counting algorithms with reasonable runtimes, such as the orbit counting (ORCA) algorithm~\cite{hovcevar2014combinatorial}.
Please refer to the recent survey of Ribeiro et al.~\cite{ribeiro2021survey} for a more detailed introduction to motif and graphlet detection algorithms.

\subsection{Network Motif Visualizations}
In visualization research, motifs are typically used 
to display static networks. 
However, in such scenarios, the concept of over-representation is not taken into account.
For instance, Dunne and Shneiderman~\cite{dunne2013motif} simplify and depict fan or clique motifs as glyph representations.
Other related static network visualizations are utilizing motif-based features to explore motif frequencies in biological networks~\cite{schreiber2005mavisto}, to cluster networks~\cite{von2009visual}, to display large signaling networks~\cite{ma2009snavi}, to visually search networks~\cite{von2010smart}, to explore biological mutation graphs~\cite{lenz2014visual}, or to structurally explore large networks~\cite{chen2018structure}.
For example, EgoNav~\cite{harrigan2012egonav} enables users to explore summarized ego-networks utilizing motif analysis and dimensionality reduction methods.
Moreover, Kwon et al.~\cite{kwon2017would} used graphlet frequencies to compute the similarity between networks.
As for network matrix visualizations, motifs are visual patterns in a matrix, such as a line pattern.
For instance, \emph{HiPiler}~\cite{lekschas2017hipiler} allows users to explore matrix snippets (motifs) in genome interaction matrices.
All of the aforementioned motif-based visualizations are for static networks.
However, the listed static network visualizations are not suitable for providing a scalable overview of changes in dynamic networks.
The work of von Landesberger et al.~\cite{von2009system} is a unique system in this category.
The proposed system allows users to aggregate user-specified motifs and highlight local motif changes by utilizing a what-if-analysis.
However, the system is not suited for analyzing changes in dynamic networks since the system only allows investigating the impact of individual differences on local motif structures. 
Although the utility of motif-based visualizations is well-known for static networks, they have not been utilized to present dynamic networks.

\subsection{Dynamic Network Visualizations}\label{sec:rw-dynamic-vis}
Previous dynamic network visualizations often display the evolving data as timeline visualizations to reduce the complexity and provide an overview of temporal changes~\cite{beck2017taxonomy}.
For example, van Elzen et al.~\cite{van2016reducing} apply dimensionality reduction methods to embed dynamic network snapshots to connected points in a 2D scatterplot.
However, such dimensionality-based abstraction methods depend on the graph size, the frequency of changes, the used distance metric, the extracted global or local metrics (e.g., node degrees), and the non-linear dimensionality reduction methods.
Moreover, van den Elzen et al.~\cite{van2013dynamic} extend \textit{Massive Sequence Views} to visually explore dynamic networks, including simple sub-network structures, such as star patterns.
However, the approach does not support visual detection of distinct motifs, such as feed-forward loops, which are of interest in gene regulator networks~\cite{shen2002network}.
Likewise, the approach does not scale well to an increasing number of nodes and networks, causing more overlap and clutter, making the visual detection of sub-networks challenging.
Hadlak et al.~\cite{hadlak2013supporting} cluster domain attributes to detect groups of nodes with similar trends and behavior.
However, the approach only clusters time-varying node and edge attributes, creating clusters without considering or utilizing the network topology.
Bach et al.~\cite{bach2014graphdiaries} propose \textit{GraphDiaries} utilizing animated transitions to navigate and highlight changes in a dynamic network. 
Yet, animations are unsuitable for displaying large-scale dynamic networks due to the high cognitive effort to compare and trace changes over time~\cite{tversky2002animation}.
Other dynamic visualization approaches utilize persistent homology~\cite{hajij2018visual} or display dynamic networks on large physical displays~\cite{lee2019dynamic}.
However, again both approaches do not allow exploring motif-based changes in dynamic networks.
Recently, Xie et al.~\cite{xie2020interactive} proposed \textit{MeasureFlow} to explore time-series of network metrics (e.g., network density) to provide an overview of changes in dynamic networks. 
The approach also enables tracking and comparing trends of user-defined sub-networks using metrics (e.g., number of connected nodes) as superimposed line and bar charts.
Yet, the approach's usefulness depends on the user-selected sub-networks, the network size, and the frequency of changes, including the selected temporal granularity. 
\textit{MeasureFlow} also does not scale to a large number of motifs since every motif requires a single line chart. 
For further readings, please refer to the surveys of Kerracher et al.~\cite{kerracher2015task} and Beck et al.~\cite{beck2017taxonomy}.

Recently, visualization researchers proposed initial pixel-based visualizations for dynamic networks.
Pixel visualizations can present large amounts of data without overlap and clutter~\cite{keim2000designing}, being dense and ultimately able to scale to large datasets.
They are generally useful, among others, for visual explorations of groups, trends, correlations, and outliers in large datasets~\cite{behrisch2018quality}.
Only a few pixel-based dynamic network visualizations have been proposed, which we present next in chronological order. 
First, Stein et al.~\cite{stein2010pixel} proposed pixel-based glyphs to present temporal patterns in an adjacency matrix. 
The proposed method works only for small social networks and does not allow motif exploration. 
Second, Burch et al.~\cite{burch2011parallel} proposed the parallel edge splatting approach to display a series of static as bipartite layouts, including the interleaving concept~\cite{burch2017visualizing} to increase the approaches scalability. 
However, the proposed approaches are only helpful for visually exploring edges and their attributes. 
Next, Cui et al.~\cite{cui2014let} proposed \emph{GraphFlow} to display structural changes of metrics in dynamic networks using a pixel and energy-based visualization.
However, the \emph{GraphFlow} method depends on the node metric (e.g., node degree) and can only display smaller networks.
Archambault and Hurley~\cite{archambault2014visualization} present a design study to highlight trends in telecommunication networks as pixel-oriented visualizations, focusing on displaying clustered privacy-preserving histogram data.
Again the approach clusters social network data based on temporal domain attributes (summary histograms), neglecting the temporal analysis of network topologies.
Recently, Cakmak et al.~\cite{cakmak2020dg2pix} proposed \emph{dg2pix}, a multiscale pixel-based visualization to highlight temporal states and changes in dynamic networks. 
Yet, the approach's usefulness depends heavily on non-transparent graph embeddings, posing the challenge of mapping latent space changes to explicit structural changes.
Contrary to dg2pix~\cite{cakmak2020dg2pix}, our approach is interpretable and provides an overview of significantly occurring sub-network structures.
We thereby ensure that the visible patterns are not merely random in some latent space since we do not use non-linear dimension reduction techniques and only extract interpretable features.
In addition, our approach also allows comparing multiple dynamic networks and single networks against each other.

\subsection{Delineation to our Work}\label{sec:delineation}
We compare a selection of related work to delineate our work and highlight the research gap we intend to close in Table~\ref{tab:comparison}.
The compared dimensions comprise the following aspects: the visualization type, the scalability regarding the number of networks, the visually analyzed sub-network structures, and the sub-group analysis tasks based on the network evolution task taxonomy by Ahn et al.~\cite{ahn2013task}.
\begin{table}[tb]
  \scriptsize%
	\centering%
	\resizebox{\columnwidth}{!}{%
  \begin{tabular}{|r?c|c|c?c|c|c?c|c|c?c|c|c|c|c|c|c|}
  \hline
    \textbf{Publication, Year} &
    \multicolumn{3}{|c|}{\textbf{Visualization}}  &
     \multicolumn{3}{|c|}{\textbf{Scalability}} & 
    \multicolumn{3}{|c|}{\textbf{Sub-Network}} & 
    \multicolumn{5}{|c|}{\textbf{Temporal Sub-Group Tasks}} \\ \specialrule{.1em}{.05em}{.05em} 
    
    & \rotatebox{90}{Static} 
    & \rotatebox{90}{Animation} 
    & \rotatebox{90}{Timeline}
    & \rotatebox{90}{Small }
    & \rotatebox{90}{Medium }
    & \rotatebox{90}{Large }
    & \rotatebox{90}{Sub-Networks} 
    & \rotatebox{90}{Motifs}
    & \rotatebox{90}{Graphlets} 
    & \rotatebox{90}{Structural Properties \vspace{0.1cm}} 
    & \rotatebox{90}{Domain Attributes}  
    & \rotatebox{90}{Individual Features}   
    & \rotatebox{90}{Shape of Changes} 
    & \rotatebox{90}{Rate of Changes}
    \\ \specialrule{.1em}{.05em}{.05em} 

    \rowcolor{row3!25} \textbf{Our Approach} 
    & $\bullet$ & & $\bullet$ & $\bullet$ & $\bullet$ & $\bullet$ & $\bullet$ & $\bullet$ & $\bullet$ & $\bullet$ & & $\bullet$ & $\bullet$ & $\bullet$  \\  \specialrule{.1em}{.05em}{.05em} 
    \rowcolor{row1!25} Schreiber et al.~\cite{schreiber2005mavisto}, 2005 & $\bullet$ & &  & & &  & & $\bullet$ &  & & & & & \\ \hline
    \rowcolor{row1!25} von Landesberger et al.~\cite{von2009visual}, 2009 & $\bullet$ & & & & &  & $\bullet$ & $\bullet$ &  & $\bullet$ & & $\bullet$ & & \\ \hline
    \rowcolor{row1!25} Ma'ayan et al.~\cite{ma2009snavi}, 2009 & $\bullet$ & &  & & &  & $\bullet$ & $\bullet$ &  & & & & & \\ \hline
    \rowcolor{row1!25} von Landesberger et al.~\cite{von2010smart} 2010 & $\bullet$ & &  & & &  & $\bullet$ & $\bullet$ & & & & & & \\ \hline
    \rowcolor{row1!25} Harrigan et al.~\cite{harrigan2012egonav}, 2012 & $\bullet$ & &  & & &  & $\bullet$ & $\bullet$ &  & & & & & \\ \hline
    \rowcolor{row1!25} Dunne et al.~\cite{dunne2013motif}, 2013 & $\bullet$ & &  & & &  & $\bullet$ & $\bullet$ &  & & & & & \\ \hline
    \rowcolor{row1!25} Lenz et al.~\cite{lenz2014visual}, 2013 & $\bullet$ & &  & & &  & $\bullet$ & $\bullet$ &  & & & & & \\ \hline
    \rowcolor{row1!25} Lekschas et al.~\cite{lekschas2017hipiler}, 2017 & $\bullet$ & & & & &  & $\bullet$ & $\bullet$ & & & & & & \\ \hline
    \rowcolor{row1!25} Kwon et al.~\cite{kwon2017would}, 2017 & $\bullet$ & & & & &  & $\bullet$ & & $\bullet$ & & & & & \\ \hline
    \rowcolor{row1!25} Chen et al.~\cite{chen2018structure}, 2018 & $\bullet$ & & & & &  & $\bullet$ & $\bullet$ & $\bullet$ & & & & & \\ \hline
    \rowcolor{row2!25} Bach et al.~\cite{bach2014graphdiaries}, 2014 & & $\bullet$ &  & $\bullet$ & &  & $\bullet$ & & & $\bullet$ & & $\bullet$ & & \\ \hline
    
    \rowcolor{row2!25} Stein et al.~\cite{stein2010pixel}, 2010 & & & $\bullet$ & $\bullet$ & &  & & & & & & & & \\ \hline
    \rowcolor{row2!25} Burch et al.~\cite{burch2011parallel}, 2011 & & & $\bullet$ & $\bullet$ & $\bullet$ &  & & & & & & & & \\ \hline
    \rowcolor{row2!25} Cui et al.~\cite{cui2014let}, 2014 & & & $\bullet$ & $\bullet$ & $\bullet$ &  & & & & & & & & \\ \hline
    \rowcolor{row2!25} van den Elzen et al.~\cite{van2016reducing}, 2016  & & & $\bullet$ & $\bullet$ & $\bullet$ & & & & & & & & & \\ \hline
    \rowcolor{row2!25} Lee et al.~\cite{lee2019dynamic}, 2019 & & & $\bullet$ & $\bullet$ & $\bullet$ & & & & & & & & & \\ \hline
    \rowcolor{row2!25} Hajij et al.~\cite{hajij2018visual}, 2018 & & & $\bullet$ & $\bullet$ & $\bullet$ & $\bullet$ & & & & & & & & \\ \hline
    \rowcolor{row2!25} Cakmak et al.~\cite{cakmak2020dg2pix}, 2020 & & & $\bullet$ & $\bullet$ & $\bullet$ & $\bullet$ & & & & & & & & \\ \hline
    
    \rowcolor{row2!25} van den Elzen et al.~\cite{van2013dynamic}, 2013 & & & $\bullet$  & $\bullet$ & &  & $\bullet$ & &  & $\bullet$ & & $\bullet$ & & \\ \hline
    \rowcolor{row2!25} Hadlak et al.~\cite{hadlak2013supporting}, 2013 & & & $\bullet$  & $\bullet$ & $\bullet$ &  & $\bullet$ & &  & & $\bullet$ & & $\bullet$ & $\bullet$ \\ \hline
    \rowcolor{row2!25} Archambault et al.~\cite{archambault2014visualization}, 2014 & & & $\bullet$ & $\bullet$ & $\bullet$ &  $\bullet$ & $\bullet$ & & & & $\bullet$ & & $\bullet$ & $\bullet$ \\ \hline
    
    \rowcolor{row2!25} Xie et al.~\cite{xie2020interactive}, 2020 & $\bullet$ & & $\bullet$ & $\bullet$ & $\bullet$ & $\bullet$ & $\bullet$ & $\bullet$ & & $\bullet$ & & $\bullet$ & $\bullet$ & $\bullet$ \\ \hline

  \end{tabular}%
}
 \vspace{0.1cm}
 \caption{
    The comparison delineates our approach from related work. 
    The table is colored by static (blue) and dynamic network (orange) visualizations, with an additional sub-ordering by visualization and sub-network type. 
    The first category \textbf{visualization} shows if the work utilizes static network visualizations, animation, or timeline representations based on the taxonomy of Beck et al.~\cite{beck2017taxonomy}.
    The \textbf{scalability} category indicates the temporal scalability with small ($<100$), medium ($<1000$), and large ($>1000$) number of networks in a dynamic network.
    The \textbf{sub-network} category reveals the visually analyzed structures.
    The last category describes the supported \textbf{temporal sub-group tasks} for network evolution analysis~\cite{ahn2013task}.
    }
    \vspace{-2\baselineskip}
    \label{tab:comparison}
\end{table}

\begin{figure*}[tb]
 \centering 
 \includegraphics[width=0.94\linewidth]{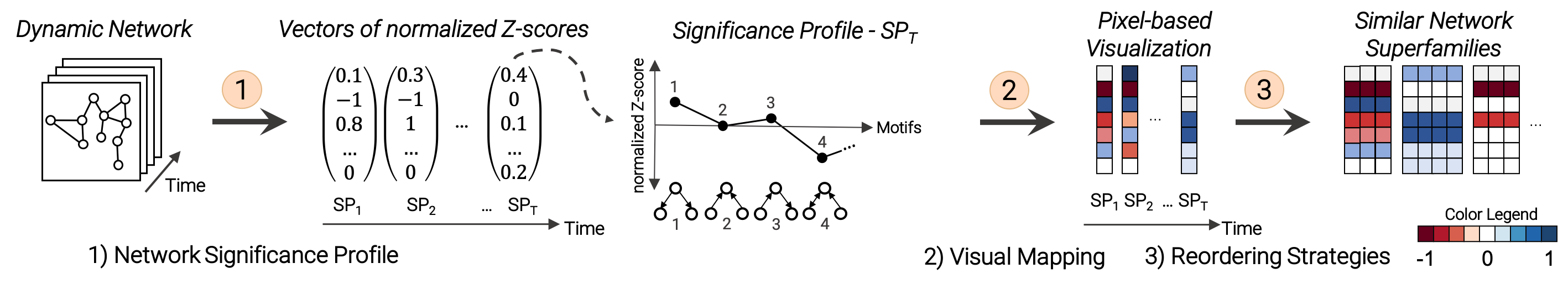}
 \vspace{-1\baselineskip}
 \caption{
 The pipeline displays the steps for generating the network-level census visualization.
 (1) the network motif significance profile (census) is calculated for each time step, (2) the vectors are presented as a pixel-based visualization, and (3) reordering strategies are used to reveal similar network superfamilies.
 $SP_T$ illustrates the relation of the vector values to their respective motifs.
 The reordering strategies are crucial for grouping similar network topologies to emphasize structural changes, trends, states, and outliers. 
 }
 \vspace{-1\baselineskip}
 \label{fig:pipeline}
\end{figure*}

Table~\ref{tab:comparison} reveals common features and outlines the following research gaps:
First, the publications colored in blue (see Table~\ref{tab:comparison}) utilize sub-networks to abstract and visualize static networks.
The static network visualizations are used to abstract and increase the readability of node-link diagrams and highlight common motifs.
However, all the listed static motif-based network visualizations do not allow visualizing changes within sub-networks in dynamic networks.
Second, animations of dynamic networks enable exploring structural properties and individual features. 
However, animations are not used to display sub-networks in dynamic networks since they tend to increase cognitive load for users, making it difficult to detect and trace structural changes over time~\cite{tversky2002animation}. 
Moreover, multiple dynamic network approaches (orange) provide an overview of evolving dynamic network changes without enabling the exploration of temporal sub-group tasks. 
Finally, four timeline visualizations by van den Elzen et al.~\cite{van2013dynamic}, Hadlak et al.~\cite{hadlak2013supporting}, Archambault et al.~\cite{archambault2014visualization}, and Xie et al.~\cite{xie2020interactive} allow users to explore basic group structures in dynamic networks.
However, the four papers either focus on clustering domain-specific attributes or only enable to explore simple motifs or clustered sub-network, such as star motifs in \textit{Massive Sequence Views}~\cite{van2013dynamic}.
For a more detailed delineation of the last four papers, please refer to the previous Section~\ref{sec:rw-dynamic-vis}.

In summary, to this day, dynamic network visualizations did not use motifs and graphlets, although such sub-network structures can provide useful, scalable overviews of evolving sub-network topologies.
To the best of our knowledge, our proposed approach is the first dynamic network visualization that provides an overview of evolving sub-network structures in dynamic networks.

\section{Structure-Based Visual Abstraction} \label{sec:main_section}
In this work, we propose two pixel-based visualizations: a network-level census view presenting an entire dynamic network and a detailed node-level sub-network metric view to investigate the local node neighborhoods of single networks.
The views combine motif-based network analysis with pixel-based visualizations to reveal evolving topological structures in dynamic networks and examine network topologies in detail.
Our central idea is to identify significantly occurring motifs and then analyze the motifs using scalable and clutter-free pixel visualizations.
Next, we describe the employed dynamic network model and both pixel-based visualizations.

\subsection{Dynamic Network Model}
The input to our approach is a discrete series of directed networks, such as daily snapshots of an evolving social network (see Section~\ref{sec:use-case-2}).
From a practical viewpoint, often dynamic networks are modeled as a sequence of events, such as varying connectivity between nodes.
For such cases, temporal discretization can be applied by computing supergraphs to generate static networks~\cite{hadlak2013supporting}.
However, identifying a proper temporal discretization for dynamic networks remains challenging since it depends on the application domain, the user task at hand, and the underlying evolving data.
For instance, a low temporal discretization results in a large set of static networks with no differences in motif structures. 
On the other hand, a too coarse temporal discretization leads to large static networks that may hide motif changes.
Thus, selecting a potential temporal discretization scale needs to be predefined by the user, considering that identifying a proper temporal analysis scale in dynamic networks is a non-trivial task~\cite{devineni2017one}.

\subsection{Network-Level Census Visualization}
Analyzing dynamic networks requires obtaining an overview of the diversity of structural changes in the evolving data, such as identifying changes, trends, states, and outliers ~\cite{van2016reducing}. 
Providing such an overview usually goes beyond solely counting nodes and edges for each time step. 
There is a great interest in understanding how the underlying topological structures changed.
For instance, in social networks studying the formation of triadic closures is of great interest~\cite{romero2010directed}.

Our network-level census visualization provides an overview of the evolving structural properties and reveals structural changes, trends, states, and outliers in dynamic networks.
Our visual representation enables identifying similar network structures, such as networks that consist over-proportionally of triad motifs (see Section~\ref{sec:evaluation}).
Figure~\ref{fig:pipeline} outlines the three main steps: (1) the computation of a network census (significance profile) for a set of motifs, (2) the visual mapping, and (3) the exploration of the resulting pixel-based visualization. 
Our idea is to capture significantly occurring network motifs over time and explore the resulting network census as dense overlap and clutter-free representations.
The basic idea of the computation of a network census is to reduce and abstract the number of occurring motifs in a single network into a network census vector.
The census helps identify similar networks and network superfamilies, enabling the comparison of different sized networks~\cite{milo2004superfamilies}.
Network superfamilies are groups (clusters) of networks with similar censuses and thus similar underlying network topology~\cite{milo2004superfamilies}.
In our work, we also propose to utilize by default the thirteen triad motifs without self-loops, which are basic building blocks of networks~\cite{milo2004superfamilies}.
However, the motif selection depends on the application domain and thus needs to be manually adapted based on the user task. 
For example, users might calculate induced or non-induced quads motifs census~\cite{ortmann2017efficient}.

In the second step, the vectors are displayed as a pixel-based visualization to provide an overview of the significantly evolving network structures.
The vectors are visualized as pixel bars that encode each value $SP_i$ as a colored rectangle.
We utilize a divergent colorblind-safe color scheme from ColorBrewer~\cite{harrower2003colorbrewer} to emphasize anti-motifs (red) and motifs (blue).
The colorblind-safe color scheme utilizes perceptually linear color coding for the ranges between red (under-represented), white (as expected), and blue (over-represented).
The used colors are easily distinguishable and have an intense contrast.
The network census x-axis displays, by default, the temporal dimension, and the y-axis the motifs of interest.
In the third step, we utilize different reordering and aggregation strategies to highlight visual patterns along both dimensions (see Section~\ref{sec:visual-analysis}). 

\subsection{Node-Level Sub-Network Metric Visualization}
A further challenge in dynamic networks is the in-depth comparison of networks and their topological structures at given temporal states. 
Therefore, we propose a node-level pixel visualization to compare multiple networks and nodes in more detail, using graphlets instead of network censuses.
The graphlet degree vectors (GDV) are sub-network metrics that describe the local network structure around a given node and independently of a given null model~\cite{hayes2013graphlet}. 
We utilize and display the GDVs of one network as pixel-based visualizations to investigate and compare the structural properties of individual networks in more detail. 
The main difference from the previous network-level census visualization is that we obtain one visualization for each network as the GDVs are computed for each node. 
The node-level pixel visualizations are useful for comparing the structural neighborhood of nodes in a network and several networks against each other.
For instance, visualizing such graphlets can be used to compare and align topologies of biological networks~\cite{kuchaiev2010topological}.
The visualization displays on the x-axis the nodes of the selected network and the y-axis displays the individual GDVs.
We also want to emphasize that two graphlet-based pixel visualizations of two different-sized networks will also have varying lengths. 
We use a linear grayscale color scale from ColorBrewer~\cite{harrower2003colorbrewer} for the graphlet-based visualization to highlight occurrences of local neighborhood graphlets.
The color scale highlights frequently occurring graphlets, enabling a simple comparison of GDVs.

\section{Motif-Based Visual Analysis}\label{sec:visual-analysis}
In the following, we describe the implemented prototype, which is available at the following online repository~\footnote{ \url{https://github.com/eren-ck/motif-pixel-vis}}.
The prototype consists of four central components (see Figure~\ref{fig:prototype}): A \textit{toolbar}, the \textit{network-level census view}, the \textit{node-level metric views}, and the juxtaposed \textit{network view} which displays the underlying network structure as a node-link diagram.
As for the \textit{network view}, we compute a supergraph and derive a  ForceAtlas2~\cite{jacomy2014forceatlas2} layout for the whole dynamic network to preserve the user's mental map.
Moreover, to increase the network view's scalability, we cluster networks with more than 100 nodes using the
Clauset-Newman-Moore algorithm~\cite{clauset2004finding}.
Thereby, we break down the exploration of large networks into smaller components, focusing on the existing motifs in each cluster.
Users can explore all views through zooming and panning using linking and brushing to study the exact pixel values, nodes, and edges attributes via mouseover tooltips.

\subsection{Reordering Strategies}\label{sec:reordering-strategies}
The pixel-based visualizations enable users to obtain an overview of dynamic networks through the visual analysis of similar and different pixel bars (see Section~\ref{sec:interpretation}).
However, such visual patterns may remain hidden and difficult to detect in pixel visualizations due to the vast amount of visualized data. 
Therefore, we propose several reordering strategies to reveal similar pixel bars, such as clustering network censuses to uncover similar network superfamilies.
We want to emphasize that we cannot suggest an optimal reordering strategy, considering that the reordering depends on the user task at hand, such as identifying the \textit{shape and rate of changes} in group structures as described by Ahn et al.~\cite{ahn2013task}.
Each pixel-based visualization can be seen as a $m \times n $ matrix $A$ in which each element $a_{i,j} \in  \mathbb{R}$ with $0<i<m$ and $0<j<n$. 
The matrix rows $a_{i,:}$ represent the network census values over time for the network-level census visualization and the number of graphlets for the sub-network view. 
The columns of the matrix $a_{:,j}$ encode the motif census or the graphlet degree vectors. 
Moreover, a matrix reordering is a bijective function $\varphi\to \mathbb{N}$ that maps the rows or columns with a unique new index position.  

We enable users to arrange the columns $a_{:,j}$, hence, the x-axis of the network-level and node-level views using clustering and sorting.
The clustering of the network-level census view allows for examining superfamilies of similar network topologies.
The clustering utilizes the cosine similarity between the $a_{:,j}$ vectors and HDBSCAN~\cite{campello2013density} to identify superfamilies of similar dynamic network structures.
We use by default the widely applied cosine similarity, however, other distance functions like Euclidean or Earth Mover distance can interchangeably be used.
Moreover, we utilize HDBSCAN~\cite{campello2013density}, as this approach implements a heuristic over different parameters to discover clusters with differing densities.
The vectors $a_{:,j}$ in each identified cluster are ordered by the temporal dimension.
We also allow reordering the network-level census columns $a_{:,j}$ by sorting the networks using evolving graph metrics, such as the number of edges or the average clustering coefficient of each network $N_i$.
The reordering using such metrics enables us to relate global network metrics with the evolving structural properties over time.
The columns $a_{:,j}$ of node-level metric views can be analogously reordered by clustering and sorted by node metrics, such as the page rank or centrality of a node to highlight important nodes.
In addition, we enable users to reorder the rows $a_{i,:}$ by computing statistical measures, such as the mean, minimum, maximum, variance, and standard deviation of the $SP_i$ over time and $GDV_i$ values.
The reordering of the rows lets us rank dimensions according to statistical measurements to highlight patterns, such as block and band patterns as described by Behrisch et al.~\cite{behrisch2016matrix}.

The reordering strategies help investigate changes in the underlying evolving sub-networks and provide an overview of a dynamic network.
However, pixel-based visualizations are hard to understand due to the cognitive effort to derive patterns from several thousand or more pixels simultaneously~\cite{keim2000designing}.
Therefore, we also propose aggregation methods to abstract visual patterns and reduce cognitive efforts for users.
\subsection{Aggregation Strategies}
Moreover, we cluster the vectors and allow users to expand specific clusters to analyze them in detail if the x-axis does not scale with the number of time points or nodes.
We utilize the HDBSCAN~\cite{campello2013density} algorithm to cluster the vectors, including also the temporal aspect for the network-level census view.
We do not cluster the y-axis since we expect both visualizations to scale up to 1000 motifs.
We include the temporal aspect into the clustering process for the network-level census view by adjusting the similarity metric using a temporal filtering threshold $\varepsilon_{time}$. 
We compute the distance matrix between all network censuses using the cosine similarity. 
Afterward, we filter the distance matrix using an epsilon $\varepsilon_{time}$ for the temporal dimension to cluster only temporally close time steps.
For example, for dynamic networks with a daily temporal granularity,  $\varepsilon_{time}=7$ filters and detects clusters of network censuses that lie within a week interval.
The $\varepsilon_{time}$ value can be set in the interface and is per default set to ten to consider only temporary close networks.

\begin{figure}[tb]
 \centering 
 \includegraphics[width=\linewidth]{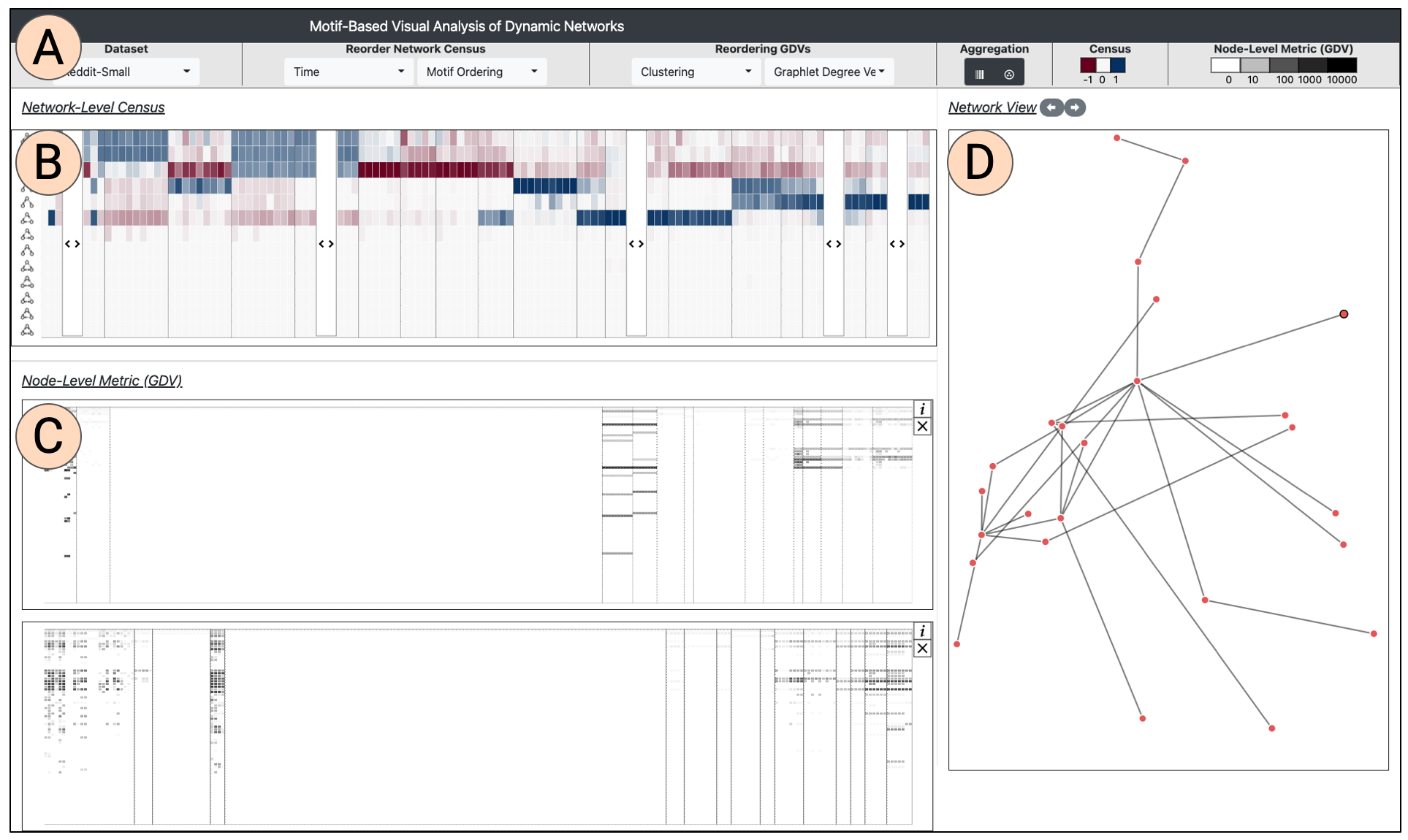}
 \vspace{-1\baselineskip}
 \caption{
    The prototype consists of four central components: (A) the toolbar, (B) the network-level census view, (C) the node-level sub-metric views, and (D) the network view.
 }
 \vspace{-1\baselineskip}
 \label{fig:prototype}
\end{figure}

For the second sub-network metric visualizations, we are utilizing the standard HDBSCAN~\cite{campello2013density} algorithm using the cosine similarity between the graphlet degree vectors. 
In both pixel-based visualizations, an exploration of the found clusters is possible by expanding them.
We display the collapsed cluster visualization as abstracted versions of each cluster.
The abstracted version depicts the first three and the last three vectors, plus an unfold button which indicates hidden vectors (see Figure~\ref{fig:prototype}-(B)).
Overall, the aggregation increases the visual scalability of both pixel visualizations.
Naturally, users can combine reordering and aggregation strategies to emphasize and reveal visual patterns in each cluster. 

\section{Visual Patterns}\label{sec:interpretation}
\begin{figure}[tb]
 \centering 
 \includegraphics[width=\linewidth]{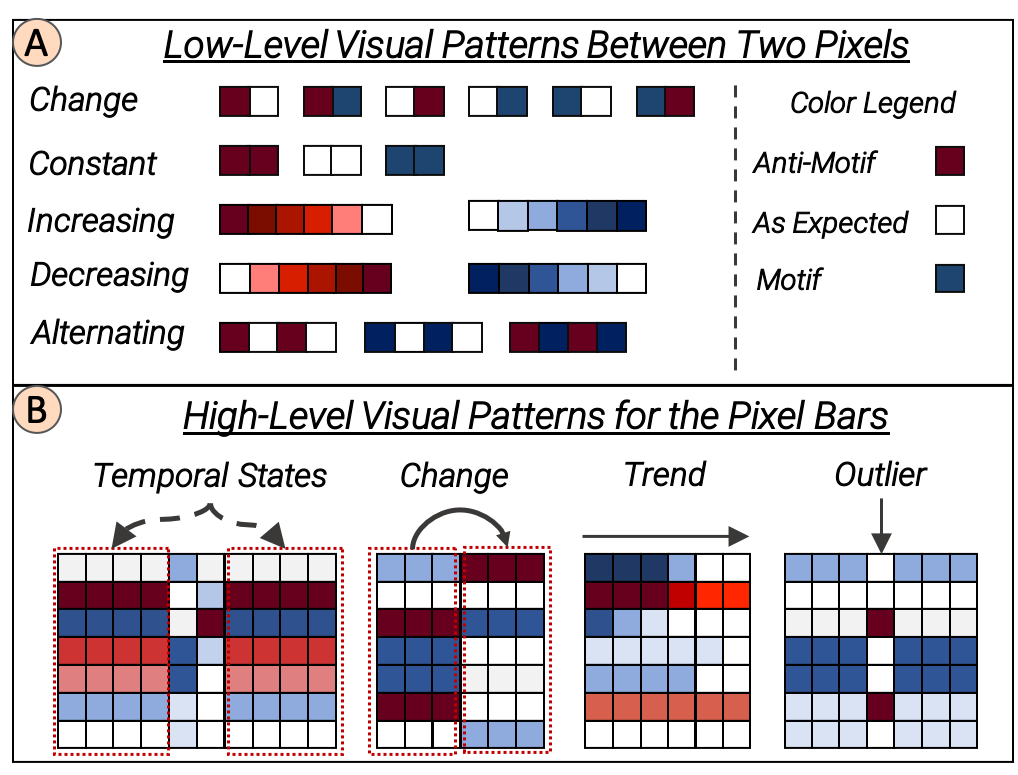}
 \vspace{-1\baselineskip}
 \caption{
    The visual patterns and their interpretations are described in Section~\ref{sec:interpretation}. 
    (A) highlights the changes between pixels in a row, and (B) displays the potential visual patterns for pixel bars.
 }
 \vspace{-1\baselineskip}
 \label{fig:visual-patterns}
\end{figure}

We want to describe the potential visual patterns for one motif in the network-level census visualization, meaning row-based changes in a series of pixels (see Figure~\ref{fig:visual-patterns}-(A)).
There are five fundamental low-level patterns in a series of colored pixels: the value \textit{changes}, remains \textit{constant}, \textit{increases} or \textit{decreases} slowly, and \textit{alternating} colored pixels.
The pattern interpretation depends on the represented motifs and the underlying dynamic network.
However, we can interpret color changes as overall shifts in the underlying networks.
For instance, a white to blue pixel color change reveals that the network topology changed, meaning that the motif now appears significantly more often than expected in a random network.  
Moreover, we expect domain-specific motifs to occur more often, such as constant anti-motifs in some social networks.
For instance, Figure~\ref{fig:use-case-2-1}-Facebook consists of chain response anti-motifs (3-triad).
In addition, we expect some visual patterns to be rare in real dynamic networks since real-world data usually does not radically change within a single time step. 
For example, changes from anti-motifs (red) to motifs (blue) between two consecutive pixels. 
If such rare patterns occur, we support examining them in more detail to understand why they appear, utilizing our network view.

Moreover, we want to present high-level visual patterns based on the described changes between pixel bars (see Figure~\ref{fig:visual-patterns}-(B)).
Changes in such pixel bars can occur for single or multiple pixel values between two pixel bars.
In the network-level view, blocks of similar pixel bars are \textit{temporal states}, and the underlying networks are groups of similar network superfamilies. 
\textit{Changes} between such temporal states are visible distinct block patterns.
Constant changes of pixel bars indicate a temporary \textit{trend} and slowly evolving network topologies. 
Finally, \textit{outlier} networks in the dynamic network are visible distinct pixel bars enclosed by similar pixel bars. 
In the sub-network metric visualization, each pixel bar encodes the actual occurrences of motifs in the node's topological neighborhood. 
The view also allows to break down large network structures and compare multiple networks by displaying the graphlet degree vector (GDV) of a node as a pixel bar (column). 
The GDV vector interpretation is relatively straightforward.
Similar local sub-network structures have similar pixel bars and vice versa.
Hence, discovering similar motif structures in large networks requires only the pairwise comparison of similar or dissimilar GDVs.

In both pixel visualizations, users have to identify similar and dissimilar pixel bars to detect relevant visual patterns.
Overall, discovering similar pixel bars is relatively simple due to the Gestalt principles of continuity, similarity, proximity, and closure~\cite[Chapter~3]{ward2010interactive}.
The similarity and proximity principles in combination let us perceive a sequence of similar pixel bars as a block. 
Similar blocks are essentially temporal states, a sequence of similar network superfamilies in the evolving network.
Moreover, the closure principle lets us perceive reoccurring blocks of similar pixel bars as repeating temporal states.
The visual analysis of the pairwise similarity between neighboring pixel bars enables users to detect temporal changes and trends in the dynamic network. 
If the pixel bars change abruptly, this indicates that the underlying structure in the dynamic network has changed drastically.
Likewise, based on the continuity principle, constant changes in the pixel bars indicate a trend of shifting network structures.
Discovering an outlier pixel bar in a block of similar pixel bars can be seen as a local or global outlier network structure.

The described visual patterns can be matched to existing dynamic network task taxonomies.
Next, we want to briefly highlight the supported tasks based on the taxonomy for network evolution analysis by Ahn et al.~\cite{ahn2013task}.
The low-level visual patterns in the network-level census view support the tasks: the shape and rate of changes of \textit{growth \& contraction}, \textit{convergence \& divergence}, \textit{stability}, \textit{repetition}, plus the \textit{fast \& slow}, and \textit{accelerate \& decelerate} for individual structural groups.
For instance, increasing and decreasing series of pixels can be used to analyze \textit{growth \& contraction} of motifs.
The low-level visual patterns enable identifying the listed shape and rate of changes. 
Moreover, the high-level patterns enable analyzing the shape and rate of changes of multiple motifs simultaneously.
Moreover, the node-level metric view supports some \textit{individual temporal feature} tasks: \textit{examining} and \textit{comparing} structural metrics using graphlet degree vectors between two time points.
For example, we can use node-level metric views to compare the number of triads or star motifs in a network and also between networks.
We also want to highlight the \textit{individual temporal feature} tasks that are not supported.
The proposed visualizations do not allow examining or tracking entities over time, such as the \textit{birth and death} of single motifs.
The two pixel-visualizations are not suitable for identifying a single motifs appearance or disappearance. 
However, the network-level census view allows the identification of such motifs if they occur significantly more often than expected in a random network. 
Likewise, our pixel-based visualizations do not support the temporal analysis of \textit{domain attributes} of nodes, links, or motifs.

\section{Use Case}\label{sec:evaluation}
\label{sec:use-case-2}
Next, we analyze real-world dynamic networks to reveal and interpret structural changes, trends, states, and outliers. 
Moreover, we provide an overview of the structural changes and compare the evolving structural properties of three real-world dynamic networks. 

\textbf{Datasets}
Figure~\ref{fig:use-case-2-1} displays the three directed dynamic networks as network-level census views.
The presented real-world dynamic networks are publicly available in the Stanford Network Analysis Project~\cite{leskovec2014SNAP}. 
The datasets were pre-aggregated to a daily temporary granularity.
Therefore, in the following, every pixel bar corresponds to one day in one of the following datasets. 
The \textit{Facebook}~\cite{viswanath2009evolution} displays wall posts between users in the City of New Orleans with 1560 days, 45.8K nodes, and 856K edges. 
The \textit{Bitcoin OTC}~\cite{kumar2016edge} presents a who-trust-who network on the Bitcoin OTC platform with 1763 days, 5K nodes, and 35K edges.
The \textit{Reddit}~\cite{kumar2018community} encodes hyperlinks (edges) in a social network between subreddits (nodes) with 1217 days, 55K nodes, and 858K edges.
We also want to briefly describe the network evolution tasks~\cite{ahn2013task} for the real-world datasets. 
Similarly, the tasks are the exploration of \textit{shape and rate of changes} using the network-level census view and the comparison of \textit{individual temporal features} using the sub-network metric view.

\begin{figure*}[tb]
 \centering 
 \includegraphics[width=0.92\linewidth]{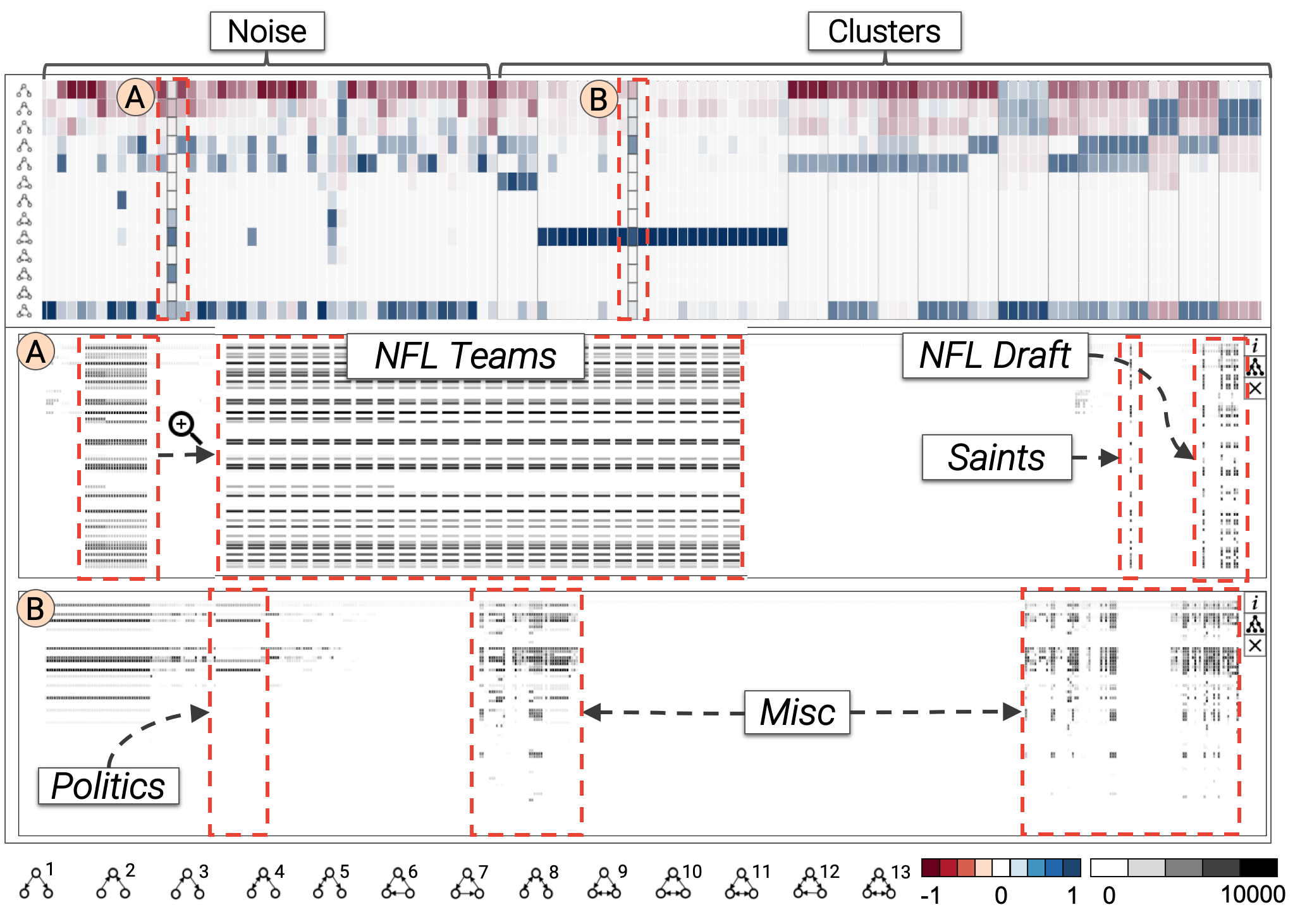}
 \vspace{-1\baselineskip}
 \caption{
    The first row displays a portion of the Reddit data~\cite{kumar2018community} as a network census view, and below that are two node-level metric views that depict two selected single networks (A-B).
    (A) highlight a distinct pixel bar in the ``noise`` cluster containing varying network censuses. 
    (B) shows an outstanding pixel bar in a cluster of similar network superfamilies. 
    The node-level metric view of (A) shows NFL subreddits being linked to trades and draft announcements.
    Moreover, (B) reveals political and misc subreddits linking each other.
    Section~\ref{sec:use-case-2} details the social network analysis.
 }
 \vspace{-1\baselineskip}
 \label{fig:use-case-2}
\end{figure*}

\textbf{Dynamic Network Exploration}
The initial striking observations are visible differences and changes in over and under-represented motifs within and across the displayed evolving networks.
The 3-triad motif representing a chain response is under-represented in all three views, visible as a low-level constant anti-motif pattern.
In particular, the 3-triad is prominently visible as an anti-motif in the Facebook dataset, meaning that chain responses on multiple Facebook walls are underrepresented. 
In this context, the first label Figure~\ref{fig:use-case-2-1}-(A) highlights a significant structural change where 4 and 5-triads started to be more frequently over-represented in May 2006, being visible as an increasing low-level visual pattern.
The 4 and 5-triads reflect the mutual posting and replying behavior between friends in the Facebook network.
The appearance of these motif triads in 2006 correlates to the growing number of Facebook users, which doubled worldwide in 2006, leading to an over-representation of the 4 and 5-triads.
The visible change in Figure~\ref{fig:use-case-2-1}-(A) is a direct result of the fact that more users joined and started to use the wall feature, creating more communication in the social network and thus motifs.

The label Figure~\ref{fig:use-case-2-1}-(B) highlights two trends in which the 8-triad is strongly over-represented on the Bitcoin OTC platform.
The trend is visible as an increasing and afterward decreasing low-level visual pattern for the 8-triad.
The 8-triad motif represents the behavior in which two users on the platform frequently give each other trust ratings after transactions.
Furthermore, the two highlighted periods can be linked to real-world events.
The first period is between May to June 2011, during which the Bitcoin price rose from \$1 to \$30, and the second period is from March to May 2013, in which the Bitcoin price briefly increased to \$250.
During these two periods, there was probably increased trading activity, so users issued more trust ratings.

Finally, Figure~\ref{fig:use-case-2-1}-(C) outlines an anomaly (outlier) period in the Reddit network between February to March 2017 in which the 9-triad is over-represented.
The visual pattern is a change, including some alternations between motif (blue) and the occurrences as expected (white). 
We can investigate the visible anomaly period by selecting the various networks with the 9-triad motif and displaying the underlying network structure as sub-network metric views.
The investigation of the detailed views reveals numerous hyperlinks between subreddits dealing with the National Football League (NFL) in the USA. 
The triads appear quite prominent following the NFL Superbowl in February 2017. 
These structural changes can be linked to real-world events. 
Between February and March 2017, there were general discussions about potential NFL player trades and draft picks, various debates about NFL players protesting during the national anthem, and a discussion about a proposed ``bathroom bill``.

\textbf{Social Network Analysis}
Next, we show how the reordering strategies and the sub-network metric view help to investigate and compare the structural properties of networks in more detail.
We continue to visually explore the Reddit hyperlink network~\cite{kumar2018community} in more detail. 
First, we reorder both axes of the network census view using clustering for the x-axis and then sort the rows of the y-axis using the median of each $SP_i$ value.
As a result, we obtain a network census view presenting clusters of network superfamilies, including a noise group.
The reordering of the y-axis also highlights block and band patterns in the pixel visualization as described by Behrisch et al.~\cite{behrisch2016matrix}.
Next, we analyze the view in more detail by zooming into specific parts of the network census view. 
The first view in Figure~\ref{fig:use-case-2} reveals distinct pixel bars in the ``noise`` group and superfamilies of similar networks in the identified ``clusters``.
The clusters are grouped temporal states previously described as high-level visual patterns. 
In each cluster, the grouped pixel bars are ordered according to time.
Furthermore, we can easily detect minor differences within the network superfamilies (clusters), which can be investigated and compared in more detail in the sub-network metric views. 
For instance, we can investigate subtle trends or outlier pixel bars within each cluster, visible as a high-level visual pattern.

\textbf{Node-Level Metric View}
Next, we provide an overview of the structural properties of single networks.
We select outstanding networks from the ``noise`` group (see Figure~\ref{fig:use-case-2}-(A)) and one prominent census from one of the clusters (see Figure~\ref{fig:use-case-2}-(B)).
The prototype then displays the two selected networks as sub-network views, which we then sort the x-axis according to node page rank in ascending order. 
Afterward, we investigate the two node-level metric views and interpret the apparent groups to expose structural differences between the two networks.
The Figure~\ref{fig:use-case-2}-(A) consists of individual nodes that are strongly connected and many individual nodes that have only one link to another node, forming the white space in the middle of the pixel visualization.
A quick exploration of the nodes via mouse-over reveals that the left group consists of NFL teams subreddits (e.g., \textit{49ers} and \textit{ravens}).
The right group consists of more general NFL subreddits (e.g., \textit{nfl} or \textit{nfl\_draft}) discussing and linking potential trades and draft announcements.
The visible similar pixel bar blocks appear after using the reordering strategy. 
We want to highlight that the NFL teams and draft groups have distinct pixel bars, meaning the nodes have different topological neighborhoods.
The NFL teams group consists of NFL teams that are all linked by the \textit{nfl} subreddit, which can be a typical bot activity to advocate trending topics in the subreddits. 
Moreover, after zooming in and analyzing the pixel bars, we can see subtle differences between the NFL team nodes, which we can then investigate in the network view. 
The subtle differences are visible as a high-level pattern of similar pixel bars and originate from the fact that some NFL team subreddits also link each other, which seems to be the typical linkage behavior of Reddit users.
In contrast, the NFL draft group consists mainly of central nodes linking to all the NFL teams. 
Moreover, there is one outlier pixel bar, the NFL team subreddit, the \textit{saints} linking to more than ten teams, indicating some bot activity again.
The second network Figure~\ref{fig:use-case-2}-(B), consists of more linked sub-network structures, including more nodes with various hyperlinks between them. 
For example, a group with political topics that link one subreddit with various nation subreddits (e.g., \textit{Sweden} and \textit{Greece}) is visible.
To the left of the political topics group, there are again numerous nodes that are linked to only one node, indicating some bot activity. 
Further, there are larger sub-networks that consist of miscellaneous topics (misc), such as computer games or education subreddits.
Again, there are sub-groups with similar pixel bars in each of the described groups, which are either disconnected or connected small sub-networks.

Finally, we want to compare the two sub-network views, revealing that both networks have different network topologies. 
In Figure~\ref{fig:use-case-2}-(A), there are many more isolated linked nodes (white space) and fewer mixed subreddits groups.
The network topology seems to be an exception in the evolving Reddit dataset.
In comparison, the network in Figure~\ref{fig:use-case-2}-B occurs more often in the dynamic network since we have a cluster of similar networks in the network census view.
The use case illustrates how we can use pixel visualizations and reordering strategies to discover similar networks and compare specific networks.

\section{Discussion}\label{sec:discussion}
Our use cases demonstrate the applicability of our approach for identifying and comparing changes, states, trends, and outliers, including superfamilies of similar sub-networks structures in large-scale dynamic networks.
Still, our approach has some limitations.

\textbf{Motifs}
We want to discuss the input parameters of our approach.
The first step in Figure~\ref{fig:pipeline} depends on the selected motifs and the used null model to compute the network census (significance profile) for each time step. 
Hence, these two parameters have to be set by a user as they heavily depend on the analyzed data properties, the application domain, and the user's task.
For instance, the null model depends on the application domain as it has to generate networks with similar topological properties to the real-world networks (e.g., similar network density).
We plan to investigate useful motifs and null model combinations in future work, including filtering motifs with particular node and edge attributes.
We propose using triad motifs as default since triads are considered the lowest level of social structures~\cite{holland1977method}, and they are used to reveal network superfamilies~\cite{milo2004superfamilies}.
Alternatively, users can select smaller or larger motifs, such as dyads or quads. 
However, discovering large motifs is computationally expensive since the runtime of motif discovery algorithms depends on the motif and network size.
Computing motif censuses for large dynamic networks are only feasible up to eight node motifs since the runtime increases dramatically starting from eight node motifs, as shown in the runtime comparison of Masoudi-Nejad et al.~\cite{masoudi2012building}.
The choice of motifs directly influences the visual patterns and, hence, the analysis and perception of changes in the dynamic network.
We consider the usage of different motifs, including motif sizes, as an advantage of our approach and a chance for further future work to examine how motif sizes affect the perceived visual patterns.

\textbf{Usability}
Pixel visualizations remain challenging to read due to the sheer amount of displayed data.
We tried to address this limitation by providing various reordering strategies to highlight similar rows and columns.
In addition, the interpretation of network census depends on the application field. 
For instance, motifs have different meanings in biology or social network analysis.
We also identified the risk that the white color pixels ($SP_i \approx 0$) might be misunderstood.
The white pixels do not necessarily imply that the motifs are not occurring in the network but rather that they are not occurring significantly more or less often than expected in a null model.
Moreover, the visible patterns depend on the used color scheme. 
For instance, the decreasing and increasing low-level visual patterns are challenging to see if the color map nuances resemble each other too much.
We plan to resolve this limitation by interactively adapting the color scheme to a user input to highlight particular visual patterns.
We want to point out that two equivalent motif censuses do not imply that the networks are identical. 
For a similar motif census, one can only conclude that the networks have a similar underlying network structure. 
Still, we cannot infer whether the network nodes or edges are identical. 
For this purpose, we propose to utilize the sub-network metric view to compare multiple networks to explore similar sub-networks and nodes.
Furthermore, the approach remains challenging for untrained users unfamiliar with network science due to the challenging interpretation of the network census, graphlet degree vectors, and the variety of proposed reordering strategies.
We also want to examine reordering strategies for different user tasks, for example, to identify helpful reordering strategies to emphasize the shape and rate of changes in network censuses. 
In future work, we plan to evaluate the approach's usability with users and develop suitable user guidance methods to analyze the pixel visualizations semi-automatically.
Moreover, the sub-network metric views are independently reordered, which means that the direct comparison of columns is currently only possible using linking and brushing. 
In future versions, we want to improve the comparison of multiple pixel visualizations through discrepancy visualizations that highlight minor differences, including exploring new reordering strategies that better align multiple sub-network metric views.

\textbf{Scalability}
The scalability of the approach poses another challenge since the computational time grows exponentially with the size of the motifs as the subgraph matching problem is known to be NP-complete~\cite{cook1971complexity}.
Overall, computing network census or applying the orbit counting algorithm is only feasibly for sub-network structures between three to eight nodes~\cite{hovcevar2014combinatorial}.
For more details regarding the execution times of various subgraph counting algorithms, please refer to the recent survey of Ribeiro et al.~\cite{ribeiro2021survey}.
In addition, the generation of null model networks can also be computationally expensive. 
We are currently generating by default 100 null models for each network census.
Therefore, we suggest precomputing the network census for each step and loading the significance profiles into the main memory.
Apart from that, the last computationally expensive aspect is the clustering of potentially large static networks for the visualizations in the network view. 
For larger networks with more than 1000 nodes, the aggregation and visualization method has to be adapted.
The survey of von Landesberger et al.~\cite{von2011visual} lists and discusses useful methods to simplify and display large networks.
In addition, the visual scalability for the network census view and the node-level metric view depends heavily on the available display space.
We increase the scalability along the x-axis by aggregating and simplifying the clusters of similar pixel bars.
However, if the clustering outputs too many clusters, this can lead to the visualization no longer scaling with respect to the x-axis.
In future work, we aim to overcome this limitation by providing an interactive multiscale clustering for the temporal dimension and also by utilizing frequent pattern analysis.
We assume that this will enable users to change the temporal and network granularity to reduce the complexity of pixel visualizations. 
Moreover, the y-axis scales up adequately for each pixel visualization up to 1000 motifs.  
Although, we expect displaying 1000 motifs is not feasible in most applications since the computation is quite expensive for an entire dynamic network.

\section{Conclusion} \label{sec:conclusion}
We presented a visualization approach to provide a scalable overview of structural changes in long and large-scale dynamic networks.
The approach utilizes network censuses and graphlet degree vectors to capture and display the structural similarities between evolving network structures as pixel-based visualizations.
The pixel-based visualizations reveal similar temporal states, trends, and outliers in dynamic networks using motifs and node-level statistics. 
Moreover, the approach allows exploring abstracted dynamic network summaries searching for temporal patterns (e.g., network superfamilies) without previous knowledge about the evolving data.
Overall, the proposed visualizations enable us to display static and dynamic networks to provide an overview of the underlying evolving structural properties. 
The main idea of our approach is not limited only to motifs or graphlets and can be generalized to display other structure-based properties (e.g., evolving roles) of dynamic networks.
In future work, we plan to integrate new methods for tracing and comparing temporal motif structures and motif sub-networks, including the same set of nodes and edges over time. 
Moreover, we plan to study network metrics and their effects on the overall changes in the evolving network to semi-automatically identify essential motifs in the network census which potentially reveal the before-mentioned network metric changes. 


\acknowledgments{This work was funded by the Deutsche Forschungsgemeinschaft (DFG, German Research Foundation) under Germany's Excellence Strategy - EXC 2117 - 422037984.}

\bibliographystyle{abbrv-doi}

\bibliography{00-template.bib}
\end{document}